\documentclass[fleqn,10pt]{wlscirep}
\usepackage[utf8]{inputenc}
\usepackage[T1]{fontenc}
\usepackage{amsmath}
\usepackage{booktabs}
\usepackage[export]{adjustbox}
\usepackage{hyperref}
\usepackage{graphicx}
\usepackage{bm}
\usepackage{subcaption}

 \newcommand{\hbarhorizline}{\mathchar'26\mkern-7mu h}

\title{Accurate Prediction of the $\alpha \to \beta$ Phase Transformation Temperature in Tin via Full Anharmonic Treatment}

\author[1,2]{Petr \v{S}est\'{a}k}
\affil[1]{Institute of Physics of Materials, Czech Academy of Sciences, v. v. i., Brno, Czech Republic}
\affil[2]{Faculty of Mechanical Engineering, Brno University of Technology, Brno, Czech Republic}

\author[3]{Matous Mrovec}
\affil[3]{Interdisciplinary Centre for Advanced Materials Simulation (ICAMS), Ruhr-Universität Bochum, Germany}

\author[1,*]{Martin Fri\'{a}k}
\affil[*]{friak@ipm.cz}

\begin{abstract}
Predicting the $\alpha \to \beta$ (grey-to-white) transition temperature in tin presents a longstanding challenge for atomistic simulations, existing theoretical approaches over- or underestimating the experimental boundary (286 K) by up to several hundred Kelvin. In this work, we construct an Atomic Cluster Expansion (ACE) potential trained on density functional theory data to evaluate the finite-temperature free energies of both phases. Evaluated on the same potential energy surface, the quasi-harmonic approximation predicts a transformation temperature of 377 K, whereas full thermodynamic integration, which accounts for explicit vibrational anharmonicity, yields 288 K.  This shift directly quantifies the explicit anharmonic free energy, which is substantial for metallic $\beta$-Sn but negligible for semiconducting $\alpha$-Sn. The anisotropic anharmonicity in $\beta$-Sn is corroborated by its excess heat capacity, temperature-driven renormalization of its vibrational spectrum, and deviations of its atomic forces and displacements from the harmonic reference. Our results demonstrate that capturing full lattice anharmonicity is essential for predicting the phase stability of tin,  while the absolute transition temperature remains limited by the accuracy of the underlying 0 K energetics.

\end{abstract}
\begin{document}

\flushbottom
\maketitle

\thispagestyle{empty}

\section*{Introduction}

Tin (Sn) is a technologically important material widely utilized in engineering applications ranging from microelectronic solder joints to protective anti-corrosive coatings. A fundamental understanding of its physical and thermodynamic properties is essential for ensuring the structural integrity of tin-containing components. A property of particular interest is the allotropic $\alpha \leftrightarrow \beta$ phase transformation. In practical applications, the unintended transition from the high-temperature metallic $\beta$-Sn phase (white tin) to the low-temperature semiconducting $\alpha$-Sn phase (gray tin), commonly referred to as ``tin pest'', leads to severe structural degradation. Because this transformation is accompanied by a large volume expansion of approximately 25–30\%, it can induce structural failure and complete mechanical disintegration of the material. At ambient pressure, the equilibrium transformation temperature between the diamond cubic (DIA) $\alpha$ phase and the body-centered tetragonal (BCT) $\beta$ phase occurs at $13.2\,^\circ\text{C}$ (286 K) \cite{farrow1981growth, vnuk1984effect}. While this experimental value is well established, its accurate prediction using atomistic simulations presents a significant challenge. 

First-principles studies of the $\alpha\leftrightarrow\beta$ transition have a long history \cite{Ihm1981, Pavone_PhysRevB.57.10421, Na2010, legrain2016, Friak2022, Chen2023, Nitol2023, Mehl2021}, which exposes how sensitive the description of Sn energetics is. Because the static 0 K energy difference $\Delta E_{\alpha\beta} = E_{\beta} - E_{\alpha}$ between the two phases is only on the order of a few tens of meV/atom~\cite{legrain2016, Mehl2021, Friak2022}, the temperature dependencies of the free energy for $\alpha$- and $\beta$-Sn cross almost tangentially.  Therefore, the predicted transition temperature is very sensitive to the details of the electronic structure calculations, such as the basis type, the exchange--correlation (XC) functional, the core electron treatment, and the convergence criteria \cite{legrain2016, Mehl2021, Friak2022}.  This sensitivity is demonstrated for selected density functional theory (DFT) calculations in Table~\ref{tab:xc_functionals} and discussed in detail below.

Early DFT calculations using the local-density approximation (LDA) \cite{Ihm1981, Pavone_PhysRevB.57.10421, Houben_PhysRevB.100.075408} yielded small positive static energy differences, with $\beta$-Sn being less stable than $\alpha$-Sn by  $10 \text{--} 20$ meV/atom at 0 K. Combined with harmonic or quasi-harmonic phonon free energies, these studies reproduced the experimental transition remarkably well. Pavone et al.~\cite{Pavone_PhysRevB.57.10421} obtained
$T_{\alpha\beta}=311$ K and Houben et al.~\cite{Houben_PhysRevB.100.075408} found a crossing at 280 K, both attributing the transition to the larger vibrational entropy of $\beta$-Sn.  However, recent convergence studies~\cite{Mehl2021, Friak2022} revealed that LDA does not determine reliably even the sign of $\Delta E_{\alpha\beta}$ and that the true 0 K ground state predicted by fully converged LDA is actually the $\beta$ phase, with an energy lower by about 20 meV/atom than that of the $\alpha$ phase.

In contrast, DFT calculations using the generalized-gradient PBE functional~\cite{Perdew1996} consistently predict the correct energy order of the two phases at 0 K and yield the static difference between $39\text{--}46\,\text{meV/atom}$~\cite{Mehl2021, Friak2022, Ko2018, Nitol2023, legrain2016}. An even larger value of $\approx 70\,\text{meV/atom}$ is obtained by the meta-GGA SCAN functional~\cite{Chen2023, Haxhijaj2026}. Legrain and Manzhos~\cite{legrain2016} suggested that the PBE/SCAN values are likely overestimated due to an improper description of the valence $s$ states ($\alpha$-Sn exhibiting a larger $s \rightarrow p$ charge transfer than $\beta$-Sn) and showed that a small Hubbard $U$ correction of about 1.0-1.5 eV on the $s$ states lowers $\Delta E_{\alpha\beta}$ to $\approx 15\text{--}25$ meV/atom. However, since $U$ is an empirical parameter this correction reproduces the expected energetics rather than providing a rigorous, first-principles description. Furthermore, a recent systematic assessment of XC functionals \cite{Haxhijaj2026} confirmed the PBE/SCAN predictions and showed that the screened hybrid HSE functional gives $\Delta E_{\alpha\beta}$ of about $110$~meV/atom. Consequently, no standard electronic-structure method currently provides an unambiguously accepted first-principles value of $\Delta E_{\alpha\beta}$.

Finite-temperature effects are most often added to the static DFT calculations through the quasi-harmonic approximation (QHA), since fully anharmonic ab initio molecular dynamics (AIMD) is restricted by supercell size and simulation time limitations.  Mehl et al.~\cite{Mehl2021} showed that all common non-LDA functionals give significantly overestimated transition temperatures within QHA, yielding values in the range of approximately 350--500 K. This raises the question of whether the QHA overestimation results from a fundamental breakdown of the harmonic picture or if it is primarily a consequence of the functional-dependent error in the static 0 K energy difference~\cite{legrain2016}.  Some studies argue that lattice anharmonicity is largely secondary or negligible because the transition lies well below melting point~\cite{legrain2016, Pavone_PhysRevB.57.10421}.  These estimates, however, were obtained within 
harmonic or quasi-harmonic frameworks, or from an upper-bound approximations, none of which captures the temperature-dependent phonon renormalization of the soft $\beta$ phase. In addition, there exists strong experimental evidence of intrinsic anharmonicity in $\beta$-Sn based on the temperature dependence of the heat capacity $C_{p}$, which significantly exceeds the classical Dulong-Petit ($3R$) limit above 200 K,\cite{Khvan_2019} and almost-forbidden X-ray reflections that can arise uniquely from anharmonic atomic motions~\cite{Merisalo_1978}. Whether explicit anharmonicity contributes significantly to the free energy of $\beta$-Sn---and thus to the transformation temperature---therefore remains an open question that the existing DFT studies do not settle.

Classical interatomic potentials have been developed to access the larger length and time scales needed for direct thermodynamic sampling, but they have not resolved this issue either. Several parametrizations based on the modified embedded-atom method (MEAM) have been developed to address the mixed metallic--covalent bonding of tin~\cite{Ravelo1997, vella2017structural, Ko2018, etesami2018thermodynamics}. The pioneering MEAM model of Ravelo and Baskes~\cite{Ravelo1997} reproduced the transformation temperature, but the experimental value was used as an explicit fitting target. In addition, this MEAM parametrization incorrectly predicts the $\beta$ phase to be dynamically unstable at low temperatures. The potential of Vella et al.~\cite{vella2017structural} was optimized for the liquid and cannot describe the solid $\alpha\leftrightarrow\beta$ transformation. The MEAM models of Etesami et al.~\citeonline{etesami2018thermodynamics}---using thermodynamic integration---and that of Ko {\it et al.}~\cite{Ko2018}---using the QHA---bracket the experimental value of transition temperature from opposite sides, predicting $158\,\text{K}$ and $460\,\text{K}$, respectively. The overall scatter of these predictions reflects the limited transferability of the MEAM functional form, which cannot simultaneously reproduce the static energetics, elastic response, and vibrational properties of both phases across a wide temperature range.

A new class of machine-learning interatomic potentials (MLIPs) can deliver the required accuracy and transferability, but existing models have so far focused on other aspects of the Sn phase diagram. The hybrid EAM--neural-network (EAM-RANN) potential of Nitol et al.~\citeonline{Nitol2023} predicted the transformation temperature of $303\,\text{K}$ through melting-point and Gibbs--Helmholtz integration, but their study did not provide any phonon or quasi-harmonic analysis. The moment-tensor and neural-network potential study of Chesser et al.~\citeonline{Chesser2025} aimed at $\beta$-Sn twin boundaries and assessed neither the phase stability nor finite-temperature free energies. The deep-potential (DP) model of Chen et al.~\citeonline{Chen2023}, trained on SCAN and PBE data, was constructed to map the high-pressure and high-temperature phase diagram rather than the ambient pressure $\alpha \leftrightarrow \beta$ transition. Crucially, none of the existing MLIP studies have isolated the vibrational contribution to the ambient transformation by directly comparing the quasi-harmonic and the fully anharmonic free energies.
 
In the present work,  we address this gap by constructing an Atomic Cluster Expansion (ACE) potential~\cite{Drautz2019} fitted to first-principles data computed with the PBE functional~\cite{Perdew1996}. The ACE model provides an accurate representation of the potential energy surface of both phases while fixing the electronic reference at a single, well-defined level of theory. Using this potential we compute the free energies of $\alpha$- and $\beta$-Sn in two ways---within the quasi-harmonic approximation and through full thermodynamic integration, which accounts for explicit anharmonicity---so that the difference between the two directly quantifies the anharmonic contribution to relative phase stability. This strategy follows the methodology recently established for other Group~IV elements by Jung et al.~\cite{Jung_2023}, who showed that explicit anharmonicity can shift transition temperatures by as much as a thousand kelvin and that full free-energy integration with MLIPs reaches ab initio accuracy in cases where harmonic approximations fail. Whereas Chen et al.~\cite{Chen2023} addressed the high-pressure phase
diagram, we focus on the ambient-pressure $\alpha \leftrightarrow \beta$ transition and on the explicit role of anharmonicity in stabilizing the $\beta$ phase. Our central finding is that, even with the standard PBE functional, the $\beta$ phase is significantly anharmonic, and that this anharmonic contribution to the free energy is essential for correctly predicting the phase transformation. 

\begin{figure}[ht!]
    \centering
     \includegraphics[width=0.99\textwidth]{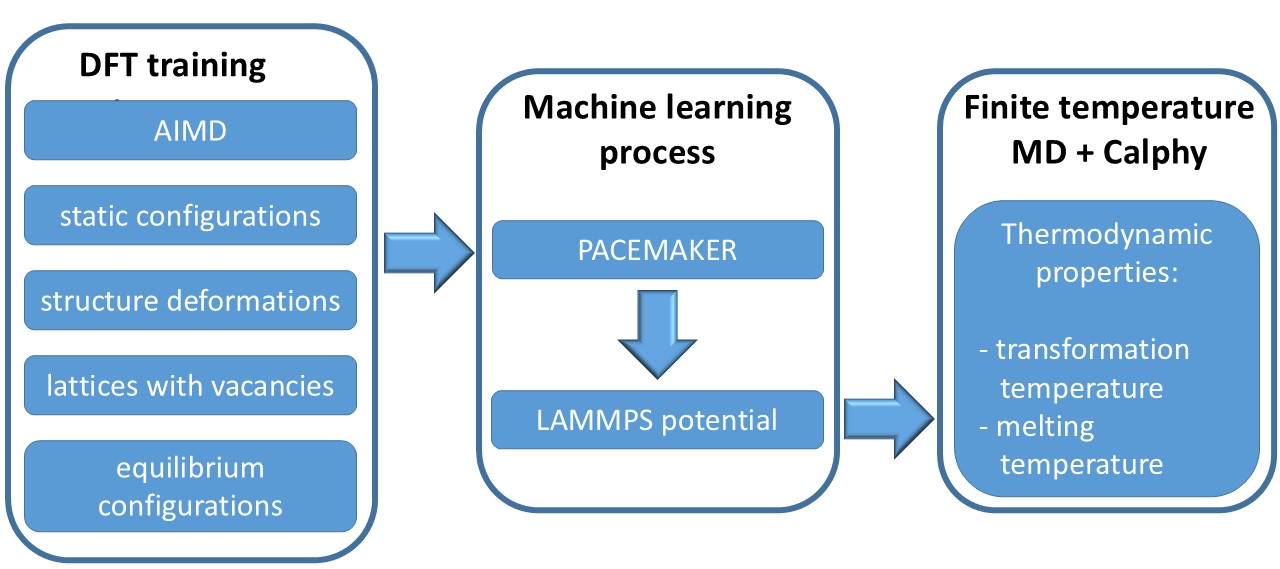}
    \caption{The basic scheme of the workflow used to predict the transformation temperature between $\alpha$-Sn  and $\beta$-Sn.}
    \label{fig:workflow}
\end{figure}

The overall workflow is summarized in Fig.~\ref{fig:workflow}: a training set of $\alpha$- and $\beta$-Sn configurations is generated using DFT calculations; an ACE potential
is fitted and used for large-scale molecular dynamics; and the Helmholtz free energies, together with the transformation temperature, are determined via thermodynamic integration.

\section*{Results}

\subsection*{Potential validation}

To validate the accuracy of the developed ACE potential, we first compare the optimized lattice parameters and the energy difference $\Delta E_{\alpha\beta}=E_\beta-E_\alpha$  with selected theoretical and experimental results in Table~\ref{tab:xc_functionals}. The ACE potential reproduces the lattice parameters of both phases and the energy difference ($\Delta E_{\alpha\beta}=+38$~meV/atom) in close agreement with the PBE DFT reference ($+42$~meV/atom).  The comparison across various DFT functionals shows the characteristic PBE/SCAN overestimation of the lattice parameters (underbinding). LDA reproduces the lattice parameters well, but ---once properly converged--- it incorrectly renders the $\beta$ phase as the 0 K ground state~\cite{Mehl2021, Friak2022}. The classical and ML potentials included in the second block reproduce their respective fitting targets.

\begin{table}[tb]
\centering
\caption{Theoretical and experimental equilibrium lattice parameters of $\alpha$-Sn (diamond, $A4$)
and $\beta$-Sn (white tin, $A5$) and the static energy difference
$\Delta E_{\alpha\beta} = E_\beta-E_\alpha$. The first block contains DFT results for different exchange-correlation functionals, the second block contains results from different interatomic potentials, the last row are experimental estimates. A positive $\Delta E_{\alpha\beta}$ means $\alpha$-Sn
is the lower-energy phase.}
\label{tab:xc_functionals}
\begin{tabular}{lccccc}
\toprule
 & $\alpha$-Sn & \multicolumn{3}{c}{$\beta$-Sn} & \\
\cmidrule(lr){2-2}\cmidrule(lr){3-5}
Method & $a$ (\AA) & $a$ (\AA) & $c$ (\AA) & $c/a$ & $\Delta E_{\alpha\beta}$ (meV/atom) \\
\midrule
PBE (this work)       & 6.65 & 5.94 & 3.22 & 0.543 & $+42$ \\
PBE \cite{Mehl2021}   & 6.65 & 5.95 & 3.21 & 0.540 & $+39$ \\
PBE \cite{Chen2023}   & 6.65 & 5.95 & 3.21 & 0.540 & $+42$ \\
LDA \cite{Mehl2021, Friak2022}   & 6.48 & 5.79 & 3.13 & 0.540 & $-23$ \\
LDA \cite{Pavone_PhysRevB.57.10421}  & 6.38 & 5.70 & 3.10 & 0.544 & $+22$ \\
LDA \cite{Houben_PhysRevB.100.075408}  & --- & 5.68 & 3.07 & 0.541 & $+12$ \\
SCAN \cite{Mehl2021}  & 6.54 & 5.89 & 3.16 & 0.536 & $+74$ \\
SCAN \cite{Chen2023}  & 6.57 & 5.91 & 3.16 & 0.536 & $+75$ \\
r$^2$SCAN \cite{Haxhijaj2026}   & 6.57 & 5.88 & 3.19 & 0.543 & $+125$ \\
HSE \cite{Haxhijaj2026}         & 6.56 & 5.86 & 3.18 & 0.542 & $+110$ \\
\midrule
ACE (this work)            & 6.64 & 5.95 & 3.21 & 0.539 & $+38$ \\
DP-PBE \cite{Chen2023}     & 6.65 & 5.95 & 3.23 & 0.545 & $+42$ \\
EAM-RANN \cite{Nitol2023}  & 6.66 & 5.92 & 3.25 & 0.548 & $+46$ \\
MEAM \cite{Ko2018}         & 6.58 & 5.86 & 3.21 & 0.547 & $+33$ \\
\midrule
Experiment        & 6.483$^{a}$ & 5.815$^{b}$ & 3.164$^{b}$ & 0.544 & --- \\
\bottomrule
\end{tabular}
\\[3pt]
{\footnotesize
$^{a}$Ref.~\cite{Price1971} (at $90\,\text{K}$).\quad
$^{b}$Ref.~\cite{Rowe1965} (at $100\,\text{K}$).\quad
\par}
\end{table}

The elastic constants are compared with DFT and literature data in Table~\ref{tab:alpha_beta_elastic}.  For $\alpha$-Sn, the ACE values agree well with the PBE reference, with only a minor deviation in $C_{12}$.  For $\beta$-Sn, the agreement is also good with the shear constants $C_{44}$ and $C_{66}$ being somewhat underestimated relative to DFT.  We note that $C_{11}$ and $C_{12}$ of $\beta$-Sn are unusually difficult to converge in DFT due to subtle ionic relaxations: despite extensive tests (energy cutoff up to $1200$~eV, dense $k$-point sampling, and varied strain magnitude), the uncertainty of the presented values is  within about $\pm5$~GPa. Nevertheless, our PBE value is consistent with the recent PBE calculation of Chen et al.~\cite{Chen2023}, whereas SCAN results deviate markedly for several moduli. It is worth noting that for $\alpha$-Sn no direct measurement of elastic constants exists. The listed experimental values are derived from inelastic-neutron-scattering phonon dispersions~\cite{Price1971, Zdetsis1977} and carry correspondingly large uncertainty. Therefore, a more meaningful benchmark for $\alpha$-Sn is the agreement of the ACE acoustic branches with the measured dispersion rather than the tabulated elastic constants. Overall, the PBE-level description and experiment agree within these uncertainties.

\begin{table}[hb]
\centering
\caption{Comparison of elastic constants $C_{ij}$ (GPa) 
for $\alpha$-Sn and $\beta$-Sn from different theoretical and experimental methods. The theoretical values correspond to $0\,\text{K}$. The experimental data for $\alpha$-Sn were estimated based on neutron scattering measurements performed at 90 K; for $\beta$-Sn, the experimental values correspond to temperatures of $4.2$ and $300\,\text{K}$ (in brackets). }
\label{tab:alpha_beta_elastic}
\begin{tabular}{l c c c c c c c c c c}
\toprule
 & \multicolumn{3}{c}{$\alpha$-Sn} & \multicolumn{6}{c}{$\beta$-Sn} \\
\cmidrule(lr){2-4} \cmidrule(lr){6-11}
Method & $C_{11}$ & $C_{12}$ & $C_{44}$ & & $C_{11}$ & $C_{12}$ & $C_{13}$ & $C_{33}$ & $C_{44}$ & $C_{66}$ \\
\midrule
ACE (this work) & 55 & 34 & 23 & & 69 & 42 & 31 & 77 & 12 & 15 \\
PBE (this work)      & 56 & 26 & 25 & & $71$ & 38 & 30 & 84 & 19 & 20 \\
PBE  \cite{Chen2023} & 55 & 26 & 43 & & 73 & 44 & 29 & 84 & 19 & 22 \\
SCAN \cite{Chen2023} & 53 & 31 & 23 & & 109 & 12 & 34 & 103 & 24 & 25 \\
Experiment$^{a, b}$  & 69-71 & 29-33 & 32-36 & &
83(72) & 58(59) & 34(36) & 103(88) & 27(22) & 28(24) \\
\bottomrule
\end{tabular}
\\[3pt]
{\footnotesize
$^{a}$For $\alpha$-Sn: Inferred from neutron scattering measurements at $90\,\text{K}$.~\cite{Price1971,Zdetsis1977} \\
$^{b}$For $\beta$-Sn: Measured from $4.2 - 300\,\text{K}$.~\cite{Rayne1961} \\
\par}
\end{table}

\begin{figure}[th!]
    \centering
    \includegraphics[width=0.995\textwidth]{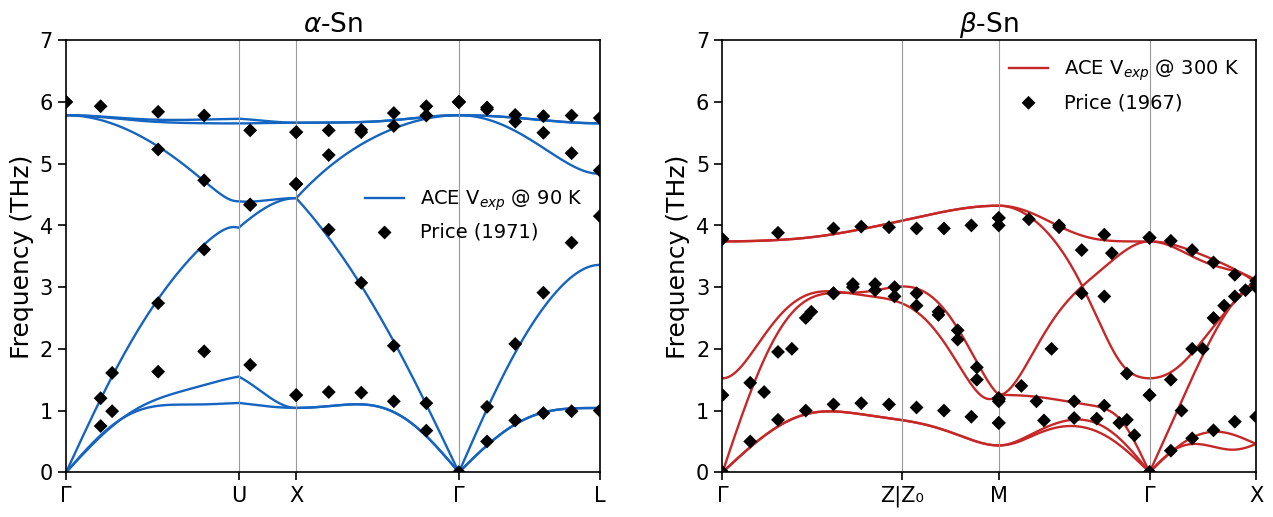} \\
    \vspace{0.5cm}
    \includegraphics[width=0.995\textwidth]{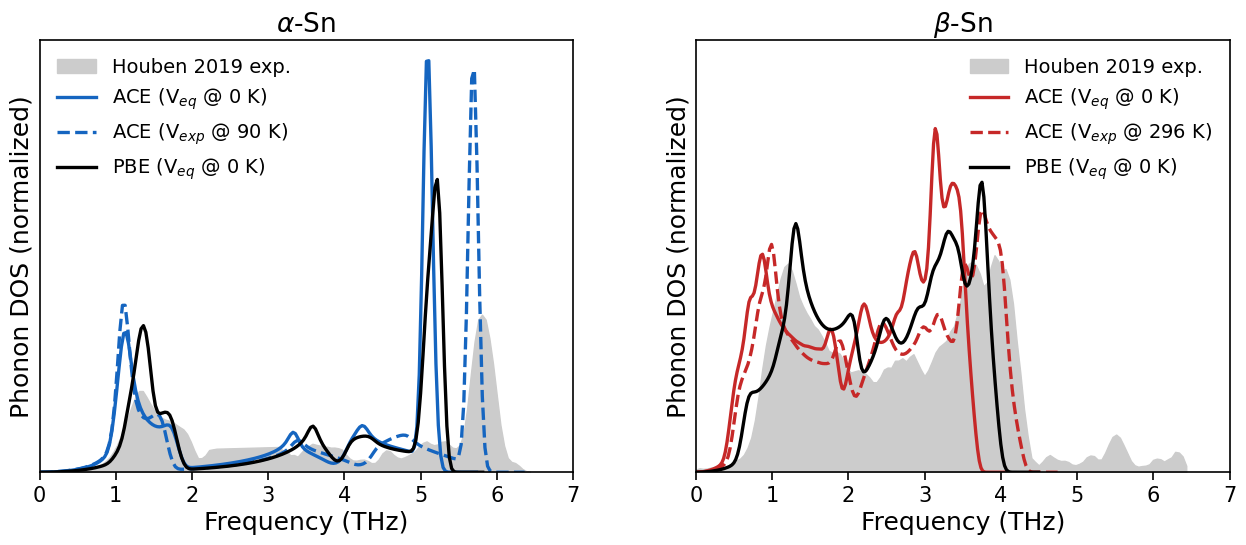}
    \caption{Comparison of phonon band structures (top row) and phonon DOS (bottom row) for $\alpha$-Sn (left column) and $\beta$-Sn (right column) from ACE calculations with available experimental data.}
    \label{fig:phonons}
\end{figure}

The phonon dispersion relations and phonon densities of states (DOS) are shown in Fig.~\ref{fig:phonons}. The ACE potential reproduces the DFT phonon spectra of both phases well. For $\alpha$-Sn, the agreement is excellent across the entire Brillouin zone. For $\beta$-Sn, small deviations are visible for the acoustic branches near the M and X zone boundaries. Crucially, the ACE spectra contain no imaginary modes, confirming that both phases are dynamically stable within the ACE description unlike some MEAM parametrizations, which render $\beta$-Sn dynamically unstable~\cite{Ko2018}. The lower panels compare the computed and experimental phonon DOS and highlight the influence of volume. Because PBE (and hence ACE) overestimate the equilibrium volume, the DOS curves computed at the corresponding equilibrium volumes are too narrow (the maximum frequencies are too low). When the DOS are evaluated at the experimental (smaller) lattice parameters, ACE results agree closely with experimental data, including the $\alpha$-Sn optical-peak position and the broader $\beta$-Sn DOS. This sensitivity of the spectra to volume already signals the importance of vibrational contributions to the phase stability.

\subsection*{QHA free energy and transformation temperature}

\begin{figure}[tb!]
    \centering
    \begin{subfigure}{0.48\textwidth}
        \centering
        \includegraphics[width=\textwidth]{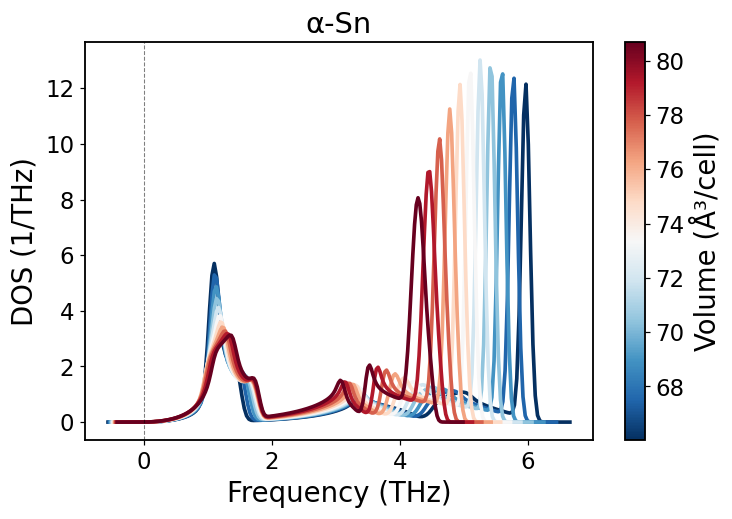}
    \end{subfigure}
    \hfill
    \begin{subfigure}{0.48\textwidth}
        \centering
        \includegraphics[width=\textwidth]{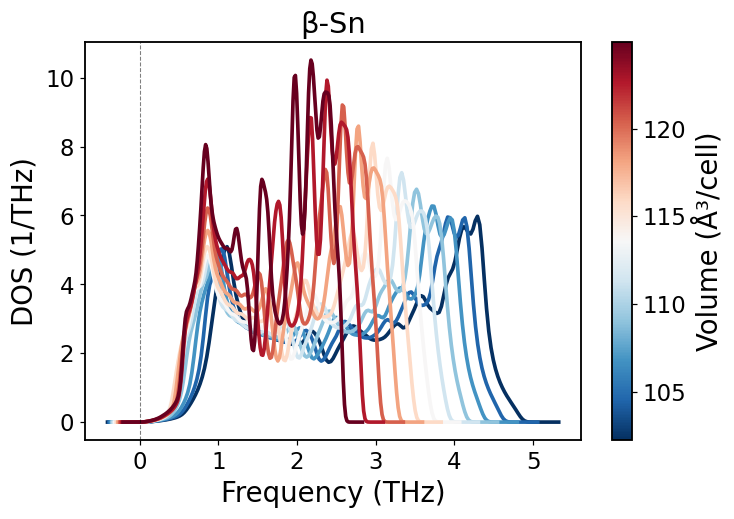}
    \end{subfigure}
    \caption{
    Variations of the phonon densities of states (DOS) for $\alpha$-Sn (left) and $\beta$-Sn (right) as functions of unit cell volumes predicted by ACE for QHA estimations of free energies. The color gradient indicates the volume per cell, with blue representing compressed and red representing expanded volumes.
    }
    \label{fig:qha_dos}
\end{figure}

The volume dependence of the phonon DOS is the basis for QHA evaluation of the free energy. Variations of DOS predicted by ACE are displayed for both phases in Fig.~\ref{fig:qha_dos}. As expected, the DOS spectra broaden/contract under volume compression/expansion, but their shapes and characteristic peaks are preserved under the volume changes sampled here.

Minimizing the QHA free energy [cf. Eq.~(1)] over volume for each phase yields the free-energy curves presented as dashed lines in Fig.~\ref{fig:transformation} (left panel). Their crossing defines the quasi-harmonic transformation temperature, $T_{\alpha\beta}^{\mathrm{QHA}}\approx377\,\mathrm{K}$. This value overestimates the experimental transition of $286$~K by nearly $90$~K, consistent with the systematic QHA overestimation reported by Ko et al.~\cite{Ko2018}, ($460\pm5$~K for the 2NN-MEAM and $470\pm5$~K for PBE DFT) and Mehl et al.~\cite{Mehl2021} ($\approx400$~K for PBE DFT). The fact that the QHA description built on several different potentials and DFT functionals consistently overshoots indicates that the overestimation is a consequence of the harmonic treatment itself rather than an artifact of any particular potential.

The reason why QHA is able to qualitatively capture the driving force of the transition is the existence of softer, lower-frequency vibrational modes in $\beta$-Sn compared to $\alpha$-Sn (cf. Fig.~\ref{fig:phonons}). This results in a larger vibrational entropy $S_{\mathrm{vib}}$ of $\beta$-Sn and the $\alpha \rightarrow \beta$ transformation, since the $-T \Delta S_{\mathrm{vib}}$ term eventually compensates the $0$~K energy preference of $\alpha$-Sn.  This entropy-driven phase transformation was already assumed in early first-principles studies~\cite{Ihm1981, Pavone_PhysRevB.57.10421, Houben_PhysRevB.100.075408}.  The QHA therefore correctly reproduces the qualitative trend, but the magnitude of the effect remains uncertain.

\subsection*{Thermodynamic integration}

\begin{figure}[tb]
    \centering
    \includegraphics[width=0.9\textwidth]{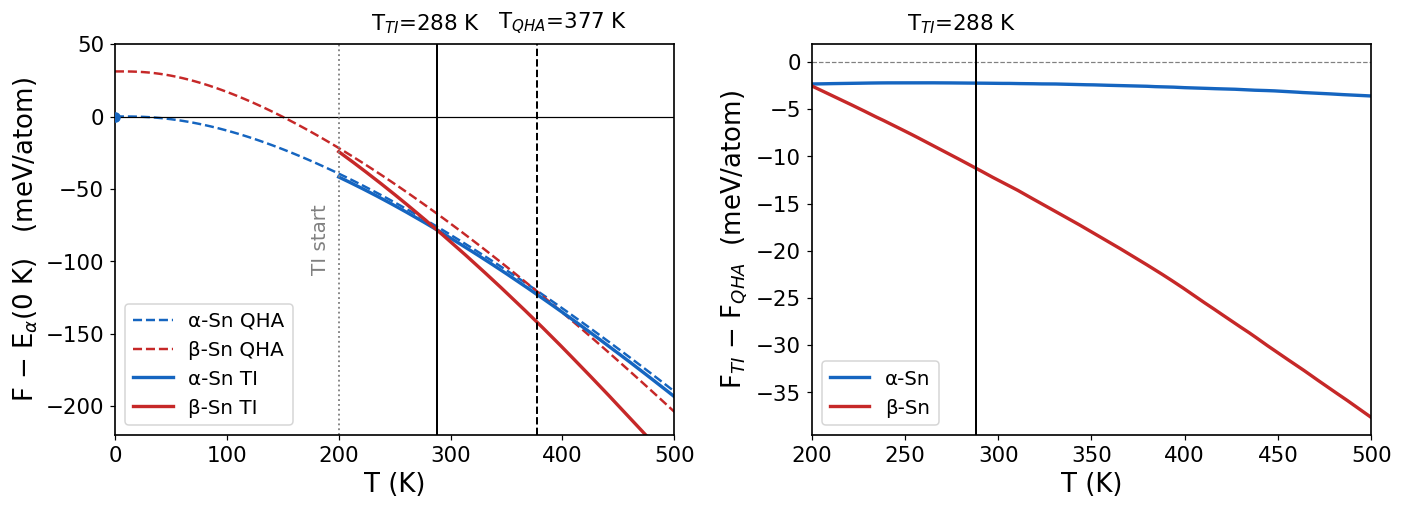}
    \caption{Temperature dependence of Helmholtz free energies $F(T)$ w.r.t. the energy of $\alpha$-Sn at 0 K  obtained via QHA and TI approaches (left panel) and the difference between TI and QHA contributions (right panel). The vertical dashed and full lines mark the $\alpha \rightarrow \beta$ transition temperatures predicted by QHA ($T_{QHA} = 377\,\mathrm{K}$) and TI ($T_{TI} = 288\,\mathrm{K}$), respectively.}
    \label{fig:transformation}
\end{figure}

The accurate determination of vibrational entropy requires full consideration of explicit anharmonic effects.  The ACE free-energy curves obtained from TI calculations are plotted as solid lines in Fig.~\ref{fig:transformation} (left panel). They cross at $T_{\alpha\beta}=288$~K, in excellent agreement with the experimental value of $286$~K. Because both QHA and TI results are derived using the same ACE potential, their difference enables to extract the explicit phonon--phonon anharmonic contribution to the free energy. The $\approx85$~K reduction from the QHA to the TI transition temperature is a direct, model-internal measure of this contribution, as illustrated in the right panel of Fig.~\ref{fig:transformation}.


This explicit anharmonic free energy for each phase is defined as $\Delta F_{\mathrm{anh}}(T)=F_{\mathrm{TI}}(T)-F_{\mathrm{QHA}}(T)$. For $\alpha$-Sn, it remains small (about $-2$ meV/atom) over the whole considered temperature range, while for $\beta$-Sn it decreases almost linearly with temperature, reaching about $-11$~meV/atom at $300$~K. The $\beta$ phase is therefore far more anharmonic than the $\alpha$ phase, and its explicit anharmonic free energy provides substantial additional stabilization. This stabilization is the quantitative origin of the QHA$\rightarrow$TI shift. 


\subsection*{Heat Capacity}

\begin{figure}[tb]
    \centering
    \includegraphics[width=0.45\textwidth]{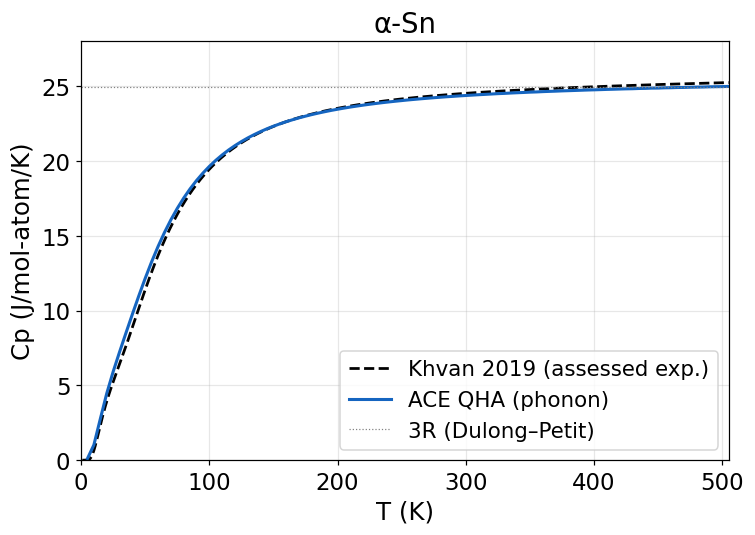}
    \includegraphics[width=0.45\textwidth]{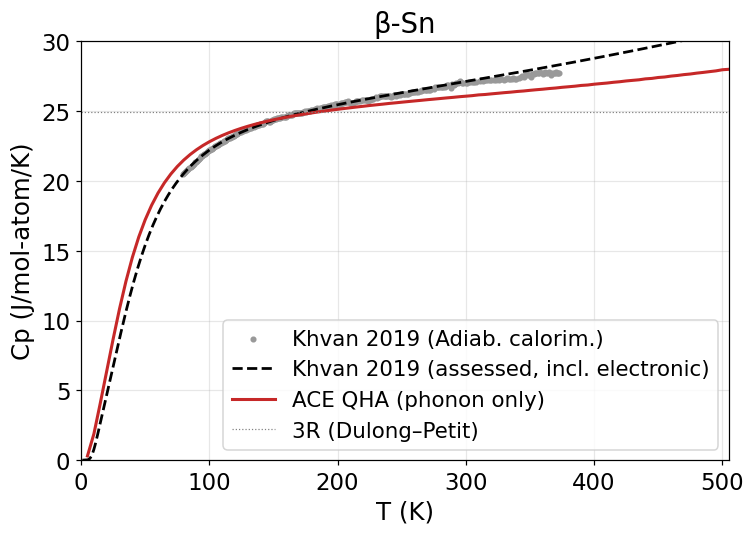}

    \caption{Calculated QHA constant-pressure heat capacity $C_p$ as a function of temperature for $\alpha$-Sn (blue line) and $\beta$-Sn (red line). The dashed horizontal line represents the classical Dulong-Petit limit of $3R \approx 24.94\,\text{J/mol}\cdot\text{K}$. The experimental data are compiled from a detailed comparative analysis by Khvan et al.~\cite{Khvan_2019}}

    \label{fig:heat_capacity}
\end{figure}

We next examine the constant-pressure heat capacity $C_p(T)$ and compare the QHA predictions with the experimental data of Khvan et al.~\cite{Khvan_2019} in Fig.~\ref{fig:heat_capacity}.  For $\alpha$-Sn, $C_p$ rises monotonically toward the Dulong--Petit limit of $3R$, while for $\beta$-Sn it exceeds $3R$ around $200$~K and keeps increasing with temperature.  The excess of $C_p$ above $3R$ originates from three contributions. The first contribution originates from the  dilation term $C_p-C_V=\alpha_V^2 B_T V_m T$, which is nonzero for any expanding solid and is therefore contained in the QHA framework. Our simulations confirm that for the soft, strongly expanding $\beta$ phase, the dilation contribution is appreciable. The second anharmonic contribution, arising from explicit phonon-phonon coupling, is captured only by the full TI and is reflected by the TI$-$QHA difference $\Delta F_{\mathrm{anh}}$ (Fig.~\ref{fig:transformation}, right panel).  Finally, since $\beta$-Sn is metallic there exists an electronic contribution $C_{\mathrm{el}}=\gamma_{\mathrm{el}}T$. However, because the electronic free energy contribution is less than $0.5$~meV/atom at 300 K~\cite{Mehl2021}, this term is negligible. 




From a thermodynamic standpoint, the $C_p$ behavior is crucial for understanding the phase stability of tin. Since the vibrational entropy is given by the integral $S_{vib} = \int (C_p/T) dT$, the higher heat capacity of the $\beta$ phase leads to a more rapid increase in its entropy as temperature increases. This entropic gain effectively lowers the Gibbs free energy of the $\beta$-phase relative to the $\alpha$ phase, thereby lowering the $\alpha \to \beta$ transition temperature. The calculated QHA heat capacity of $\alpha$-Sn is in excellent agreement with the experimental data \cite{Khvan_2019}.  For $\beta$-Sn, the QHA description recovers accurately the experimental dependence up to 200 K, but above this temperature it increases more slowly than the experimental data due to missing contributions from phonon-phonon interactions.

\subsection*{Anharmonicity analysis}

\begin{figure}[tb!]
    \centering
    \begin{subfigure}{0.48\textwidth}
        \centering
        \includegraphics[width=\textwidth]{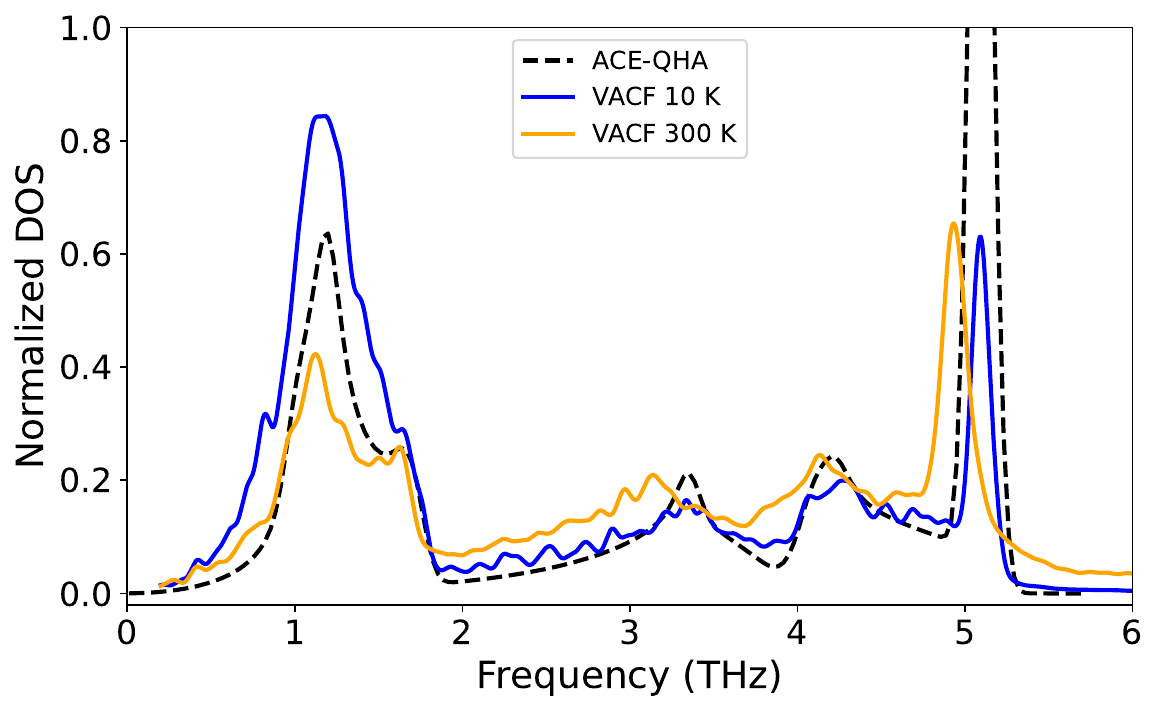}
        \caption{$\alpha$-Sn}
    \end{subfigure}
    \hfill
    \begin{subfigure}{0.48\textwidth}
        \centering
        \includegraphics[width=\textwidth]{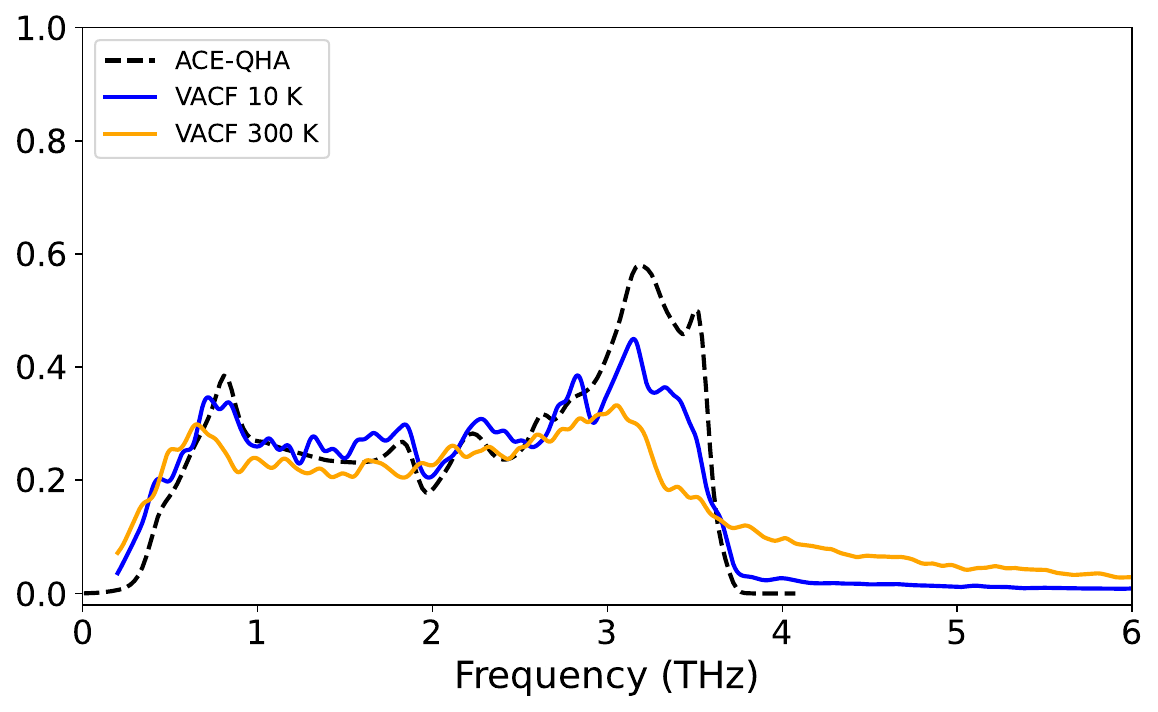}
        \caption{$\beta$-Sn}
    \end{subfigure}
    \caption{
    Comparison of VACF and QHA phonon densities of states for $\alpha$-Sn (left) and $\beta$-Sn (right) for selected temperatures.
    }
    \label{fig:vacf_dos}
\end{figure}

To further confirm and quantify the anharmonic effects at finite temperatures, we evaluated the vibrational DOS from the velocity autocorrelation function (VACF) extracted during MD simulations, as shown in Fig.~\ref{fig:vacf_dos}. 

For $\alpha$-Sn, the VACF spectrum stays close to the harmonic DOS, exhibiting only a modest transfer of weight from the high-frequency ($\sim5$~THz) peak toward lower frequencies as temperature increases---consistent with its small $\Delta F_{\mathrm{anh}}$. For $\beta$-Sn, the changes are much more pronounced.  The peak near $3$--$3.5$~THz present in the harmonic DOS is progressively suppressed with increasing temperature  and essentially vanishes above the transition temperature.  The spectral weight shifts to lower frequencies, and a weak tail develops beyond the harmonic cutoff. This  temperature-dependent variation of the $\beta$-Sn spectrum cannot be reproduced by the QHA and provides direct microscopic evidence of explicit anharmonicity.

A real-space probe of anharmonicity is obtained by analyzing instantaneous atomic displacements and interatomic forces from MD trajectories. The distribution of atomic displacements about their mean positions (see the Supplementary material) demonstrate anisotropic anharmonicity in $\beta$-Sn at 300 K for vibrations along the $[100]$ and $[010]$ crystallographic directions, whereas $\alpha$-Sn remains harmonic in all directions. Interestingly, the effective potential along these directions becomes harder for small atomic displacements and softer for larger vibrations. This real-space picture is consistent with the experimental detection of anharmonic atomic motion in white tin~\cite{Merisalo_1978} due to hardening of the potential along first nearest neighbor directions in the basal $xy$ plane.

We also compared the instantaneous atomic forces from ACE MD trajectories with harmonic forces predicted by Phonopy using second-order force constants for the identical snapshot configurations.    The deviation between the two approaches was quantified via a normalized residual force amplitude $A_F$, which represents the fraction of the total atomic force that cannot be described by the harmonic approximation ($A_F=0$ corresponding to a perfectly harmonic system). The metrics are summarized in Table~\ref{tab:anharmonicity}

\begin{table}[ht]
\centering
\caption{Comparison of harmonic and non-harmonic force and energy metrics for $\alpha$-Sn and $\beta$-Sn at 10 and 300 K. The anharmonic energy is defined as the difference between the vibrational energy obtained from molecular dynamics and the harmonic energy predicted from the Phonopy force constants evaluated at the equilibrium volume corresponding to each temperature.}
\label{tab:anharmonicity}
\begin{tabular}{llcccccc}
\toprule
Phase & $T$ (K) & $A_F$ & Force cosine & $E_{\rm harm}$ & $E_{\rm vib}$ &
$E_{\rm anh}$ & $E_{\rm anh}/E_{\rm vib}$ \\
& & & & (meV/atom) & (meV/atom) & (meV/atom) & (\%) \\
\midrule
$\alpha$-Sn & 10  & 0.048 & 0.999 & 1.29 & 1.29 & 0 & 0 \\
$\beta$-Sn  & 10  & 0.156 & 0.988 & 1.30 & 1.29 & -0.01 & -0.78 \\
\midrule
$\alpha$-Sn & 300 & 0.243 & 0.973 & 42.7 & 39.9 & -2.77 & -6.93 \\
$\beta$-Sn  & 300 & 0.580 & 0.851 & 50.2 & 42.0 & -8.27 & -19.7 \\
\bottomrule
\end{tabular}

\end{table}

\begin{figure}[ht]
\centering
\begin{subfigure}{1.0\textwidth}
    \centering
    \includegraphics[width=\textwidth]{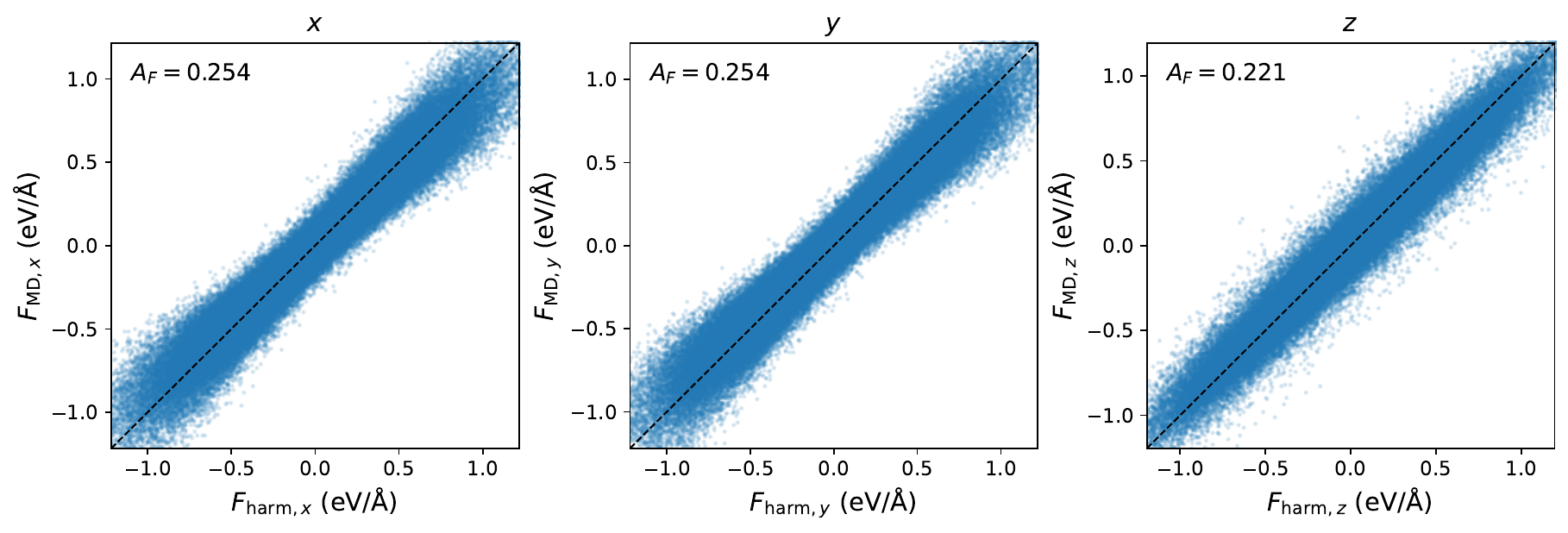}
    \caption{$\alpha$-Sn at 300 K}
    \label{fig:force_alpha_300K}
\end{subfigure}
\vspace{0.8em}
\begin{subfigure}{1.0\textwidth}
    \centering
    \includegraphics[width=\textwidth]{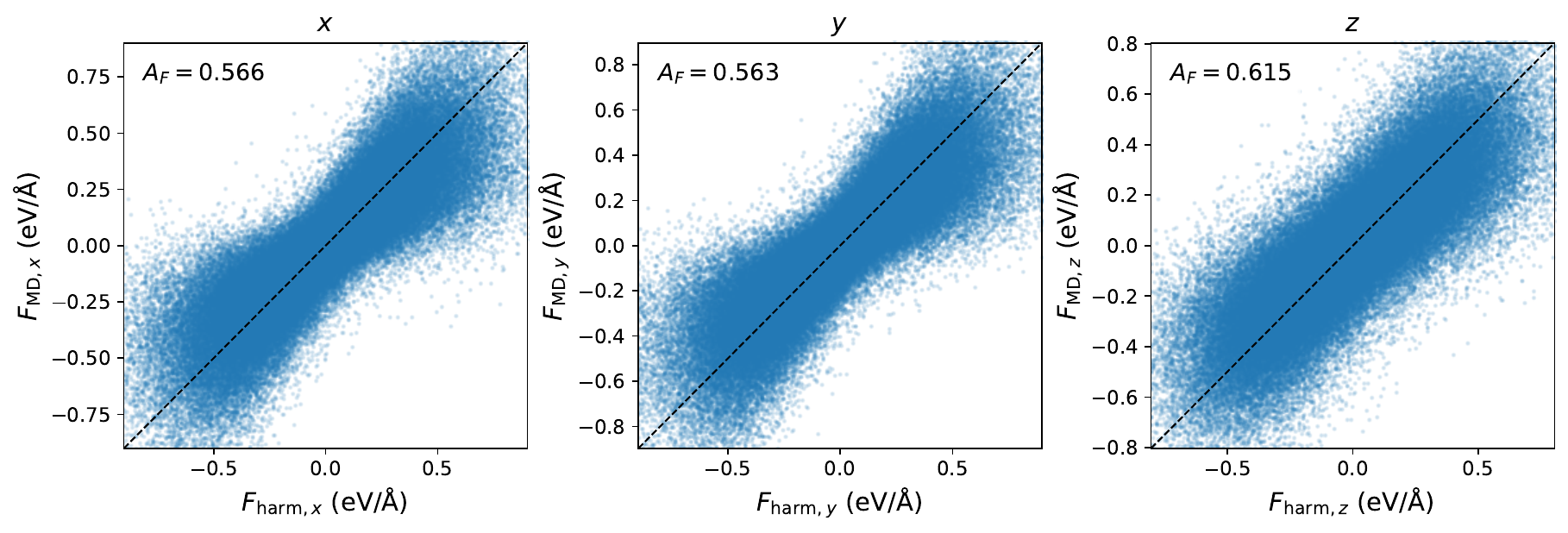}
    \caption{$\beta$-Sn at 300 K}
    \label{fig:force_beta_300K}
\end{subfigure}
\caption{Comparison between harmonic forces predicted from the second-order force constants and the instantaneous ACE MD forces. Each point corresponds to one Cartesian component of one atom in one sampled MD configuration. The dashed diagonal indicates perfect agreement between the harmonic approximation and MD. The harmonic model reproduces the forces in $\alpha$-Sn considerably better than in $\beta$-Sn, indicating substantially stronger deviations from the harmonic force field in the $\beta$ phase. Note that for $\alpha$-Sn, the x, y, and z axes are aligned with the [110], [1$\bar{1}$0], and [001] crystallographic directions, respectively.}
\label{fig:force_comparison}
\end{figure}

The force-based analysis also reveals a pronounced increase in the deviation from the harmonic approximation with increasing temperature for $\beta$-Sn.  At 300~K, the normalized residual force amplitude reaches $A_F=0.58$, whereas for $\alpha$-Sn it remains significantly lower ($A_F=0.24$). This indicates that, even after accounting for thermal expansion (i.e. using force constants evaluated at the corresponding finite-temperature volumes) the harmonic force field becomes substantially less representative of the instantaneous atomic forces in the $\beta$ phase, confirming that the residual deviation reflects explicit phonon–phonon anharmonicity rather than the quasi-harmonic volume effect.  These numerical results are directly reflected by the force correlations shown in Fig.~\ref{fig:force_comparison}. For $\alpha$-Sn, the instantaneous MD forces remain closely distributed around the harmonic prediction over the entire force range. In contrast, $\beta$-Sn exhibits a substantially broader distribution around the diagonal, indicating considerably larger deviations from the harmonic force field. 

The force cosine similarity quantifies the directional agreement between the harmonic forces predicted by the second-order force constants and the forces obtained from molecular dynamics, with a value of 1 corresponding to perfectly parallel force vectors. For $\alpha$-Sn, the force cosine remains nearly constant over the investigated temperature range and stays close to unity, indicating that the harmonic approximation accurately reproduces the force directions. In contrast, $\beta$-Sn exhibits a systematic decrease in the force cosine with increasing temperature, reflecting progressively larger deviations from harmonic behavior.



\section*{Discussion}

The results presented here aim to contribute to the long-standing debate regarding the theoretical predictions of the $\alpha \leftrightarrow \beta$ phase transition temperature in tin.  The central result of this work is the difference between two free-energy treatments obtained using the same ACE potential. Within the common QHA approach, the $\alpha\rightarrow\beta$ transformation temperature is overestimated by nearly $90$~K ($T_{\alpha\beta}^{\mathrm{QHA}}\approx 377$~K), while full thermodynamic integration yields $288$~K, within $2$~K of the experimental $286$~K. This shift is a measure of the explicit anharmonic free energy, which is relatively large for $\beta$-Sn but negligible for $\alpha$-Sn. The distinctive anharmonic character of $\beta$-Sn is corroborated by four independent observations: its excess heat capacity (Fig.~\ref{fig:heat_capacity}), the temperature-driven collapse of its $3$–$3.5$~THz vibrational feature in the velocity-autocorrelation spectra (Fig.~\ref{fig:vacf_dos}), the anisotropic, non-Gaussian atomic-displacement distributions in its basal plane (Supplemental Material), and the pronounced deviation of its instantaneous forces from the harmonic prediction (Fig.~\ref{fig:force_comparison}).

Our findings appear to contradict earlier DFT-based arguments that anharmonicity is secondary for this transition. In particular, Legrain and Manzhos\cite{legrain2016} estimated the anharmonic contribution to the relative $\alpha$–$\beta$ free energy to be only $\sim\!1$~meV/atom ($\sim\!10$~K), attributing the entire QHA overestimation to the DFT XC errors when predicting the static 0 K energy difference. Our value, $\Delta F_{\mathrm{anh}}\approx -11$~meV/atom for $\beta$-Sn at $300$~K, is more than an order of magnitude larger. The discrepancy may originate due to a short, non-equilibrium AIMD run mapped onto a fixed harmonic reference through a single global energy rescaling, a construction that cannot represent the temperature-dependent phonon renormalisation and coupling. Moreover, the $\beta$-Sn anharmonicity is evidenced independently by the measured excess heat capacity~\cite{Khvan_2019} and by the almost-forbidden X-ray reflections of white tin~\cite{Merisalo_1978}.

The absolute agreement of the transition temperature with experiment must nevertheless be interpreted with care, because the $0$~K energetics of tin is indeed uncertain. The static $\alpha$–$\beta$ difference is very small and strongly method dependent: it spans from $\approx 40$~meV/atom of PBE used here to $\approx 80$, $110$, and $125$~meV/atom for SCAN, HSE, and r$^2$SCAN, respectively\cite{Haxhijaj2026,Mehl2021,Chen2023}. No standard functional, including the screened hybrid, reproduces the low experimentally inferred value, and higher-level RPA or quantum Monte Carlo references are not yet available. Our ACE potential inherits the PBE value, which seems plausible, and the excellent agreement with experimental $T_{\alpha\beta}$ can only be achieved  when the anharmonicity of $\beta$-Sn is properly taken into account. Nevertheless, we cannot exclude that the quantitative agreement benefits in part from a cancellation between an overestimated static energy and the anharmonic contribution. 


\section*{Conclusions}

In conclusion, we have demonstrated that the long-standing uncertainty of the $\alpha \rightarrow \beta$ phase transition temperature in tin may be partly driven by the neglect of explicit lattice anharmonicity in the metallic $\beta$-Sn phase. By constructing a high-fidelity ACE machine-learning interatomic potential trained on first-principles PBE-DFT data, we accurately mapped the potential energy surface of both allotropes. Evaluating the finite-temperature thermodynamics using both the QHA and TI formalisms on the same potential energy surface reveals that explicit phonon--phonon interactions provide significant thermal stabilization of $\approx -11\,\text{meV/atom}$ to $\beta$-Sn at room temperature, whereas $\alpha$-Sn remains nearly quasi-harmonic. This explicit anharmonic contribution lowers the predicted transition temperature from $377\,\text{K}$ (QHA) to $288\,\text{K}$ (TI), achieving remarkable agreement with the experimental value ($286\,\text{K}$). This study establishes that accounting for explicit vibrational anharmonicity can be decisive for predicting phase boundaries in soft, polymorphic materials with shallow potential energy landscapes.


\section*{Methods}

\subsection*{First-principles calculations}
The training data were generated via first principles DFT calculations using the VASP code \cite{Kresse1993,Kresse1996}. We used the projector augmented wave (PAW) method \cite{Kresse1999}, and the PBE exchange-correlation functional within the generalized gradient approximation \cite{Perdew1996}. The plane-wave energy cutoff was set to 520~eV and calculations were performed using the Gaussian broadening scheme with a smearing width of 0.1~eV. The electronic self-consistency cycle was considered converged when the total energy change was below $10^{-6}$~eV; in selected cases, a smaller convergence criterion was applied to ensure numerical stability. The Sn pseudopotential was used with the semi-core $d$ electrons explicitly treated as valence states, as indicated by the \_d suffix in the POTCAR file (PAW\_PBE Sn\_d 06Sep2000). This choice improves the accuracy of the electronic structure and the total energy description, which is particularly important for the bonding characteristics of heavy elements such as tin.

\subsection*{Training dataset}
To ensure the robustness and accuracy of the ACE potential, a comprehensive training dataset was constructed, consisting of 4,427 configurations and a total of 813,487 atoms ($\approx 184$ atoms per structure). The dataset was designed to cover the relevant portion of the configurational space encompassing the equilibrium $\alpha$-Sn and $\beta$-Sn phases, as well as a selection of hypothetical and distorted structures to prevent nonphysical behavior during MD simulations.

Training data for the $\alpha$ and $\beta$ phases were generated using an identical sampling protocol to ensure consistent accuracy between both phases. To accurately map the potential energy surface (PES), the following configurations were incorporated into the training set:

\begin{itemize}
    \item Full and partial structural relaxations of primitive and supercell configurations.
    \item A wide range of volumetric and shape distortions, including hydrostatic, uniaxial, and biaxial deformations under compressive and tensile deformations.
    \item Shear strains to ensure an accurate description of the shear moduli.
    \item Defective configurations, including single and double vacancies, to capture the effects of local coordination changes.
    \item Trajectories from ab initio molecular dynamics (AIMD) simulations of pristine and defective (vacancy) supercells to incorporate finite-temperature effects.
    \item Stochastic atomic displacements to provide a dense sampling of the local potential energy surface around the equilibrium positions.
\end{itemize}

\subsection*{Potential fitting procedure}
The interatomic potential was constructed using the Pacemaker software package (version 2023.11.25) \cite{Lysogorskiy2021} \footnote{\url{https://pacemaker.readthedocs.io/}}. Pacemaker provides an automated framework for parametrizing ACE potentials, utilizing structural configurations and their corresponding total energies and forces as input \cite{Drautz2019,Lysogorskiy2023}. The fitting procedure minimizes the difference between the predicted and reference values, thereby ensuring an accurate and transferable description of the atomic interactions.

To define the local atomic environments, a cutoff radius of $r_{\mathrm{cut}} = 7.0$~\AA\ was employed, combined with an energy-based weighting policy and a $\kappa$ parameter of 0.3. The basis set was configured with 1,400 functions per element. The predictive accuracy of the resulting potential was evaluated using an independent validation set, yielding root-mean-square errors (RMSE) of 5.7~meV/atom for the energies and 33.0~meV/\AA\ for the forces.

\subsection*{Molecular dynamics simulations}
Large-scale atomistic simulations were performed using the LAMMPS (Large-scale Atomic/Molecular Massively Parallel Simulator) package \cite{Thompson2022} with the ML-PACE implementation \cite{Lysogorskiy2021}. To ensure computational efficiency for the large-scale systems studied in this work, we used the GPU-accelerated implementation of ML-PACE, powered by the Kokkos library \cite{Trott2022}. This framework enabled the efficient evaluation of the complex ACE descriptors and forces on AMD MI250X GPUs (LUMI supercomputer), significantly reducing the total wall-clock time required for the production runs.

\subsection*{Quasi-harmonic approximation}
As a reference for a full anharmonic treatment, we first evaluated the vibrational free energy for both phases within the quasi-harmonic approximation. Phonon frequencies were obtained with
the finite-displacement method as implemented in the Phonopy package~\cite{phonopy-phono3py-JPCM, phonopy-phono3py-JPSJ}, with the atomic forces on the displaced supercells evaluated using the ACE
potential.  
The phonon spectra were computed for a set of 11 volumes spanning $\pm 10\%$ around the $0$~K equilibrium volume of each phase. At each volume $V$ the Helmholtz free energy was constructed as
\begin{equation}
F(V,T) = E_{\mathrm{stat}}(V)
       + \frac{1}{2}\sum_{\mathbf{q},s}\hbarhorizline\omega_{\mathbf{q}s}(V)
       + k_{B}T\sum_{\mathbf{q},s}
         \ln\!\Big[1-\exp\!\big(-\hbarhorizline\omega_{\mathbf{q}s}(V)/k_{B}T\big)\Big],
\end{equation}
where $E_{\mathrm{stat}}(V)$ is the static ACE energy and $\omega_{\mathbf{q}s}(V)$
are the volume-dependent phonon frequencies. The equilibrium free energy at each
temperature and pressure was then obtained by minimization over volume,
$G(T,P)=\min_{V}\big[F(V,T)+PV\big]$, which at ambient pressure reduces to
$F(T)=\min_{V}F(V,T)$. The quasi-harmonic transformation temperature was
identified as the crossing of the resulting $F_{\alpha}(T)$ and $F_{\beta}(T)$
curves. 

\subsection*{Thermodynamic integration}
To determine the true transformation temperature, it is necessary to calculate the  free energy as a function of temperature for both phases. Since the free energy cannot be directly computed from standard molecular dynamics simulations, we employed the method of thermodynamic integration (TI), as implemented in the Calphy software package \cite{Menon2021}.

The TI approach computes the free energy difference between the target ACE potential, $U_{\text{ACE}}$, and a reference system with a known analytical free energy, $U_0$ (typically an Einstein crystal or a harmonic solid). A hybrid potential is constructed as a function of a coupling parameter $\lambda \in [0, 1]$:
\begin{equation}
    U(\lambda) = (1 - \lambda) U_0 + \lambda U_{\text{ACE}}
\end{equation}
The free energy difference $\Delta F$ is then obtained by integrating the ensemble average of the energy difference between the two potentials over the coupling parameter:
\begin{equation}
    \Delta F = \int_0^1 \left\langle \frac{\partial U(\lambda)}{\partial \lambda} \right\rangle_{\lambda} d\lambda = \int_0^1 \langle U_{\text{ACE}} - U_0 \rangle_{\lambda} d\lambda
\end{equation}
The Calphy framework automates the generation of the required structural configurations and the subsequent integration process. For each phase, a series of simulations were performed at various temperatures and $\lambda$ values to ensure a converged representation of the free energy surface. By calculating the free energies $F_\alpha(T)$ and $F_\beta(T)$, the transformation temperature $T_{\alpha\beta}$ is identified as the point where the free energies of the two phases intersect.

\subsection*{Velocity autocorrelation function}

The simulation protocol was designed as follows. Each system was first equilibrated for 20\,000 steps in the NPT ensemble at zero external pressure, followed by an additional 20\,000 steps in the NVE ensemble to remove residual thermostat and barostat effects. Subsequently, the simulation timestep was reset, the VACF calculation was initialized, and a production run of 3\,000\,000 steps was performed. Each simulation cell contained approximately 35\,000 atoms.

The velocity autocorrelation function was calculated using the implementation available in LAMMPS. For each time $t$, the VACF was evaluated as

\begin{equation}
C_{vv}(t)=\frac{1}{N}\sum_{i=1}^{N}
\mathbf{v}_i(t)\cdot\mathbf{v}_i(0),
\end{equation}

where $N$ is the number of atoms, $\mathbf{v}_i(0)$ is the velocity of atom $i$ stored at the beginning of the production run, and $\mathbf{v}_i(t)$ is the velocity of the same atom at time $t$. Repeating this evaluation at all timesteps gives the VACF time series,

\begin{equation}
\left\{ C_{vv}(t_n) \right\}_{n=0}^{M},
\qquad t_n=n\Delta t .
\end{equation}

The VACF time series was subsequently processed using a Python script. First, it was normalized by its initial value, and a residual long-time offset was removed. A Hann window was then applied prior to the Fourier transform in order to reduce finite-time truncation artifacts. The vibrational density of states was obtained from the Fourier transform of the processed VACF,

\begin{equation}
g(\omega) \propto
\int_0^{\infty}
C_{vv}(t)e^{-i\omega t}\,dt ,
\end{equation}

where $g(\omega)$ denotes the vibrational density of states. The resulting spectra were smoothed using a Gaussian filter for visualization purposes. Frequencies below 0.1~THz were excluded from the analysis to suppress numerical artifacts originating from finite simulation time and residual statistical noise.

\subsection*{Anharmonic and harmonic force comparison}

Another approach to assess anharmonicity is to compare the instantaneous atomic forces obtained from the ACE MD trajectory snapshots with the forces predicted for the identical structures by the Phonopy package. Therefore, we take the atomic configuration snapshots given by molecular dynamics and recalculate its forces via the following equation

\begin{equation}
\mathbf{F}^{\mathrm{harm}} = - {\Phi}\,\mathbf{u}.
\end{equation}

where ${\Phi}$ represents the matrix of the force constants and $\mathbf{u}$ the displacements from the optimized static configurations, e.g. $\mathbf{u} = \mathbf{r}_{MD} - \mathbf{r}_{0}$. This means that the harmonic force field given by Phonopy was constructed from the second-order interatomic force constants obtained by the finite-displacement method. These force constants represent the Hessian matrix of the potential energy surface evaluated at the equilibrium crystal structure at the volume corresponding to the simulation temperature $T$. These harmonic forces are subsequently compared with the instantaneous forces obtained directly from the ACE molecular dynamics simulation. 

To assess and quantify the difference between both force fields for each individual snapshot, we measure the residual force amplitude $A_{F}$ defined as

\begin{equation}
A_F= \frac 
{\sqrt{\left\langle\left|\mathbf{F}^{\mathrm{ACE}}-\mathbf{F}^{\mathrm{harm}}\right|^2\right\rangle}}
{\sqrt{\left\langle\left|\mathbf{F}^{\mathrm{ACE}}\right|^2\right\rangle}},
\end{equation}

where $\langle \cdots \rangle$ denotes averaging over all atoms and all sampled molecular dynamics configurations. The quantity $A_{F}$ therefore represents the fraction of the total atomic force that cannot be described by the harmonic approximation. i.e., a value of $A_F=0$ corresponds to an ideal harmonic system, while increasing $A_{F}$ indicates progressively larger deviations from the harmonic force field predictions.

During this evaluation process, we also determine harmonic and anharmonic energies $E_{\mathrm{harm}}$ of each structure snapshot where the harmonic term is given by

\begin{equation}
E_{\mathrm{harm}} = \frac{1}{2} \mathbf{u}^{\mathrm T} {\Phi} \mathbf{u}.
\end{equation}

Consequently, the resulting anharmonic energy contribution $E_{\mathrm{anh}}$ is determined as the energy difference between $E_{vib}^{pot}$ and $E_{\mathrm{harm}}$. 

\begin{equation}
E_{\mathrm{anh}} = (E_{\mathrm{MD}}^{\mathrm{pot}} - E_{0}(V_{T})) - E_{\mathrm{harm}}.
\end{equation}

where $E_{0}(V_{T})$ represents the potential energy of the fully relaxed reference structure at the volume $V_{T}$ corresponding to the temperature $T$ and $E_{\mathrm{MD}}^{\mathrm{pot}}$ is the potential energy of the instantaneous MD configuration.


\bibliography{references.bib}

\section*{Acknowledgements}
MF acknowledges financial support from the Czech Science Foundation (Project No. 22-05801S) as well as the project No. CZ.02.01.01/00/22\_008/0004631 "Materials and technologies for sustainable development" funded by the European Union and the state budget of the Czech Republic within the Jan Amos Komensky Operational Program. 
P\v{S} and MF also acknowledge support from the Czech Academy of Sciences ({\it Praemium Academiae} and the Strategy AV21, in particular the program "AI: Artificial Intelligence for Science and Society"). We acknowledge VSB – Technical University of Ostrava, IT4Innovations National Supercomputing Center, Czech Republic, for awarding this project access to the LUMI supercomputer, owned by the EuroHPC Joint Undertaking, hosted by CSC (Finland) and the LUMI consortium through the Ministry of Education, Youth and Sports of the Czech Republic through the e-INFRA CZ (grant ID: 90254).
the MetaCentrum and CERIT-SC. Access to CESNET storage facilities provided by the project e-INFRA CZ under the program "Projects of Large Research, Development and Innovations Infrastructures" (LM2018140) is appreciated. 
MM acknowledges support from the German Science Foundation (DFG Projektnummer 535647705).

\section*{Author contributions statement}

P\v{S} and MF conceived the study, MF and MM secured funding; P\v{S} and MM contributed to the ACE potential development, performed the atomistic simulations and thermodynamic integration calculations; P\v{S}, MM, and MF analyzed the results and wrote the manuscript. All authors reviewed and approved the final manuscript.


\section*{Additional information}

\section*{Data availability}
Selected input and output files: https://doi.org/10.5281/zenodo.21038903


\section*{Competing interests} 
The authors declare no conflict of interest.


\end{document}